\documentclass[twocolumn,showpacs,preprintnumbers,amsmath,amssymb,prl]{revtex4-1}
\usepackage{graphicx}
\usepackage{dcolumn} 
\usepackage{bm} 
\usepackage{longtable}
\usepackage{color} 
\usepackage[normalem]{ulem} 
\usepackage{amssymb} 
\graphicspath{{./}}

\begin{document}
\title{Spin polarization and attosecond time delay\\in photoemission from spin degenerate states of solids}
\author{Mauro Fanciulli$^{1,2}$, Henrieta Volfov\'{a}$^{3}$, Stefan Muff$^{1,2}$, J\"{u}rgen Braun$^{3}$, Hubert Ebert$^{3}$, Jan Min\'{a}r$^{3,4}$, Ulrich Heinzmann$^{5}$, J. Hugo Dil$^{1,2}$}
\affiliation{$^{1}$Institut de Physique, \'Ecole Polytechnique F\'ed\'erale de Lausanne, CH-1015 Lausanne, Switzerland
\\
$^{2}$Swiss Light Source, Paul Scherrer Institut, CH-5232 Villigen, Switzerland
\\
$^{3}$Department of Chemistry, Ludwig Maximillian University, D-81377 Munich, Germany
\\
$^{4}$ New Technologies-Research Center, University of West Bohemia, CZ-30614 Pilsen, Czech Republic
\\
$^{5}$Faculty of Physics, University of Bielefeld, D-33501 Bielefeld, Germany}
\date{\today}
\begin{abstract}
After photon absorption, electrons from a dispersive band of a solid require a finite time in the photoemission process before being photoemitted as free particles, in line with recent attosecond-resolved photoemission experiments. According to the Eisenbud-Wigner-Smith model, the time delay is due to a phase shift of different transitions that occur in the process. Such a phase shift is also at the origin of the angular dependent spin polarization of the photoelectron beam, observable in spin degenerate systems without angular momentum transfer by the incident photon. We propose a semi-quantitative model which permits to relate spin and time scales in photoemission from condensed matter targets and to better understand spin- and angle-resolved photoemission spectroscopy (SARPES) experiments on spin degenerate systems. We also present the first experimental determination by SARPES of this time delay in a dispersive band, which is found to be greater than 26~as for electrons emitted from the \textit{sp}-bulk band of the model system Cu(111).
\end{abstract}
\maketitle
Photoemission has been at the core of condensed matter physics studies for more than one century, both as a basic physical process and as platform for spectroscopic techniques \cite{Damascelli:2004}. Whereas enormous advances in the understanding of kinematics and energetics of the process have been achieved, other aspects such as the spin polarization of the photoelectrons and the time scale are yet to be explored in more detail. For instance, the low-efficiency measurement of the spin polarization has restricted the focus mainly on spin polarized electronic states, as in ferromagnets or spin-momentum locked systems, even though several interesting spin interference processes can take place in photoemission from spin degenerate states \cite{Heinzmann:2012}. As for the time scale, the common assumption is an instantaneous excitation and emission of the electron. However, recent measurements of finite time delays between electrons photoemitted from different electronic bands give results in the attosecond domain \cite{Cavalieri:2007, Neppl:2012, Neppl:2015, Locher:2015, Lucchini:2015}.

A complete description of the photoemission process in condensed matter must include the combination of all the relevant transitions occurring after photon absorption. In atoms, for example, an electron from an initial state with orbital quantum number $\ell$ is allowed by selection rules to make the two degenerate transitions $\ell\rightarrow\ell\pm1$. Whereas the orbital quantum number alone is not sufficient to fully describe the transition of electrons from a dispersive band in a solid, one can still consider (at least) two different transitions corresponding to different single-group symmetry spatial parts of the particular double-group symmetry that describes the initial state \cite{Yu:1998}. The interference of the complex matrix elements describing each transition determines the final state photoelectron wavefunction, whose phase term correspond to the phase shift between the transitions.

\begin{figure} [ht]
	\centering
		\includegraphics[width=0.45\textwidth]{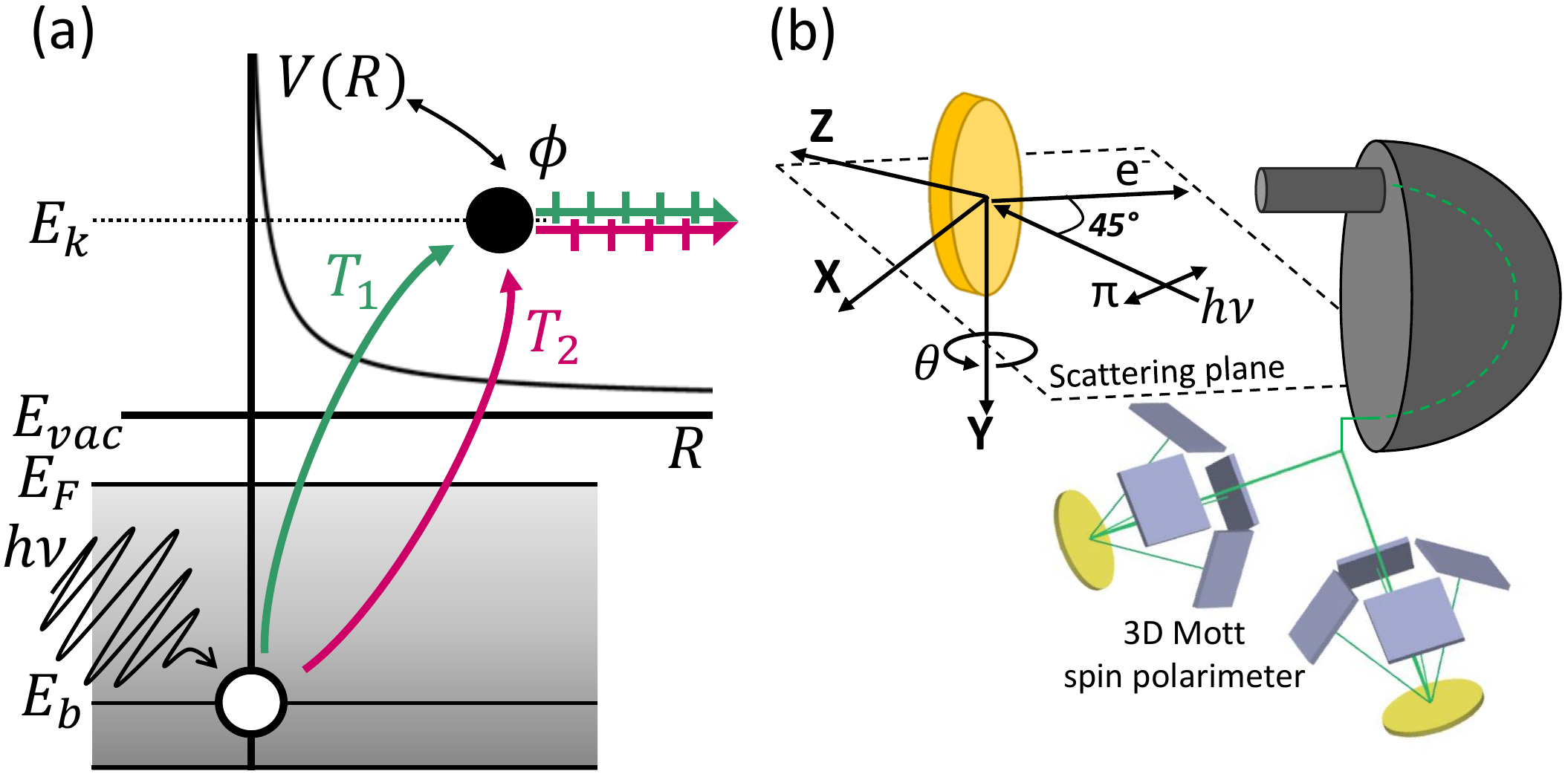}
	\caption{(a) Two phase-shifted transitions interfere in the photoemission process to build up the photoelectron wavefunction; (b) experimental setup with relevant geometry.}
	\label{fig:picture}
\end{figure}
We consider two generic transitions, $T_1$ and $T_2$, with a respective phase shift $\phi$, as pictured in Fig.~\ref{fig:picture}(a).
In the frame of the Eisenbud-Wigner-Smith (EWS) model \cite{Wigner:1955, Smith:1960}, such a phase shift corresponds to a time delay between the two transitions according to:
\begin{equation}
\tau_{\textsc{ews}}=\hbar \frac{\partial \phi}{\partial E_k}
\label{eq:EWS}
\end{equation}
where the derivative is taken with respect to the kinetic energy of the photoelectron. The original EWS model was developed for particles scattered by a short-range potential, where the phase shift is between the incoming and outcoming particles, and $\tau_{\textsc{ews}}$ is interpreted as a ``sticking'' time in the interaction potential region. A logarithmic correction to the phase shift can then be introduced to deal with Coulomb-like long-range potentials \cite{Dollard:1964}. In the photoemission context, the excited photoelectron is considered to be ``half-scattered'' by its surrounding potential \cite{Pazourek:2013}. The physical meaning of $\tau_{\textsc{ews}}$ will be discussed later.

The model proposed here permits us to exploit the capabilities of spin- and angle-resolved photoemission spectroscopy (SARPES) on non-magnetic systems \cite{Dil:2009R} in order to access the phase information and thus extract the photoemission time delay in a solid by measuring the spin polarization of photoelectrons emitted from a spin degenerate initial state maintaining full energy and angular resolution. Using the Cu(111) \textit{sp} bulk band as a model system we will present the first experimental determination of an estimate for the time delay via the spin polarization in a dispersive band.\\

In photoionization of atoms by means of linearly polarized light the photoelectrons show a spin polarization $\bm{P}=P\bm{\hat{n}}$ because of the interference between the two transitions $T_{1,2}$, as derived from ``half-scattering'' formalism \cite{Kessler:1969, Kessler:1985, Schonhense:1980}.
In particular, the spin polarization is a function of the angle $\gamma$ between the electric field of the incident light and the photoelectron momentum and is aligned along $\bm{\hat{n}}$, the perpendicular to the scattering plane defined by these two vectors \cite{Kessler:1985}. At fixed geometry $P$ is related to the atomic levels considered and to two dynamical parameters that are a function of the ratio $r=R_2/R_1$ of the radial part of the matrix elements $M_{1,2}=R_{1,2}e^{i\phi_{1,2}}$ of the two transitions $T_{1,2}$ and of their phase shift $\phi=\phi_2-\phi_1$~\cite{Huang:1980, Heinzmann:1980II, Kessler:1985, Cherepkov:1983}. In the more complicated case of molecules, it has been calculated that the direction $\bm{\hat{n}}$ can change since the symmetry of the molecule becomes relevant as well \cite{Cherepkov:1987}. In solids, it has been discussed how the spin polarization orientation is indicative of the symmetry of the bands probed by the $E$ field \cite{Schneider:1989, Tamura:1987, Schmiedeskamp:1988, Yu:1998}. The degree of polarization is evaluated as \cite{Henk:1994}:
\begin{equation}
P=I_{tot}^{-1}(\bm{\Omega})f(\bm{\Omega})\operatorname{Im}[M_1 \cdot M_2^*](r,\phi)
\label{eq:pol}
\end{equation}
where $f$ is a geometrical correction that takes into account all the relevant angles $\bm{\Omega}$, and $I_{tot}$ is the photoemission total intensity \cite{SOM}. The symmetry of the probed bands here determines $\bm{\hat{n}}$.
In photoionization of atoms the kinetic energy $E_k$ of the electron can only be changed by changing the photon energy $h\nu$. In solid state targets this implies a change in momentum $k_z$ of the probed dispersive state. However, a major advantage of dispersive states is that $E_k$ can also be varied by changing the binding energy $E_b$ for \textit{fixed} $h\nu$.

In order to access the time delay we calculate the derivative of the measured $P$ with respect to the binding energy (indicated by a dot) over the band considered and multiply by $\hbar$:
\begin{equation}
\hbar\dot{P}=\hbar\frac{\partial P}{\partial r}\dot{r}+\hbar\frac{\partial P}{\partial\phi}\dot{\phi}
\label{eq:derivative}
\end{equation}
where we do not consider the term $\partial P/\partial \bm{\Omega}$ since it is negligibly small \cite{SOM}. This equation shows that a variation of $P$ with $E_b$ is due to a time delay ($\dot{\phi}$, according to eq.~\eqref{eq:EWS}), but also in general to a change of matrix elements ratio within the band ($\dot{r}$). By rearranging eq.~\eqref{eq:derivative}, the time delay is given by:
\begin{equation}
\tau_{\textsc{ews}}=\frac{-\hbar}{\partial P/\partial\phi}(\dot{P}-\dot{r}\partial P/\partial r)
\label{eq:tau}
\end{equation}
In absence of a general theory for spin polarization due to interference in spin-degenerate states in solids, we assume that within a small $E_b$ range the double group symmetry does not vary along a given reciprocal space direction ($\dot{r}\approx 0$). This assumption is supported \textit{a posteriori} by the fact that $I_{tot}$ does not sensibly vary within the $E_b$ range considered. This means that only the phase shift of the matrix elements will vary, so that by measuring $\dot{P}$ we can estimate a finite time delay:
\begin{equation}
\left|\tau_{\textsc{ews}}\right|>c\left|\dot{P}\right|
\label{eq:approxtau}
\end{equation}
with $c=\hbar/\max{\left|\partial P/\partial \phi\right|}$. The evaluation of the coefficient $c$, the influence of additional spurious effects, an estimate of the upper limit for $\left|\tau_{\textsc{ews}}\right|$ and the analysis of corrections to the estimate when $\dot{r}\neq0$ are reported in \cite{SOM}.\\

\begin{figure*} [ht]
	\centering
		\includegraphics[width=0.95\textwidth]{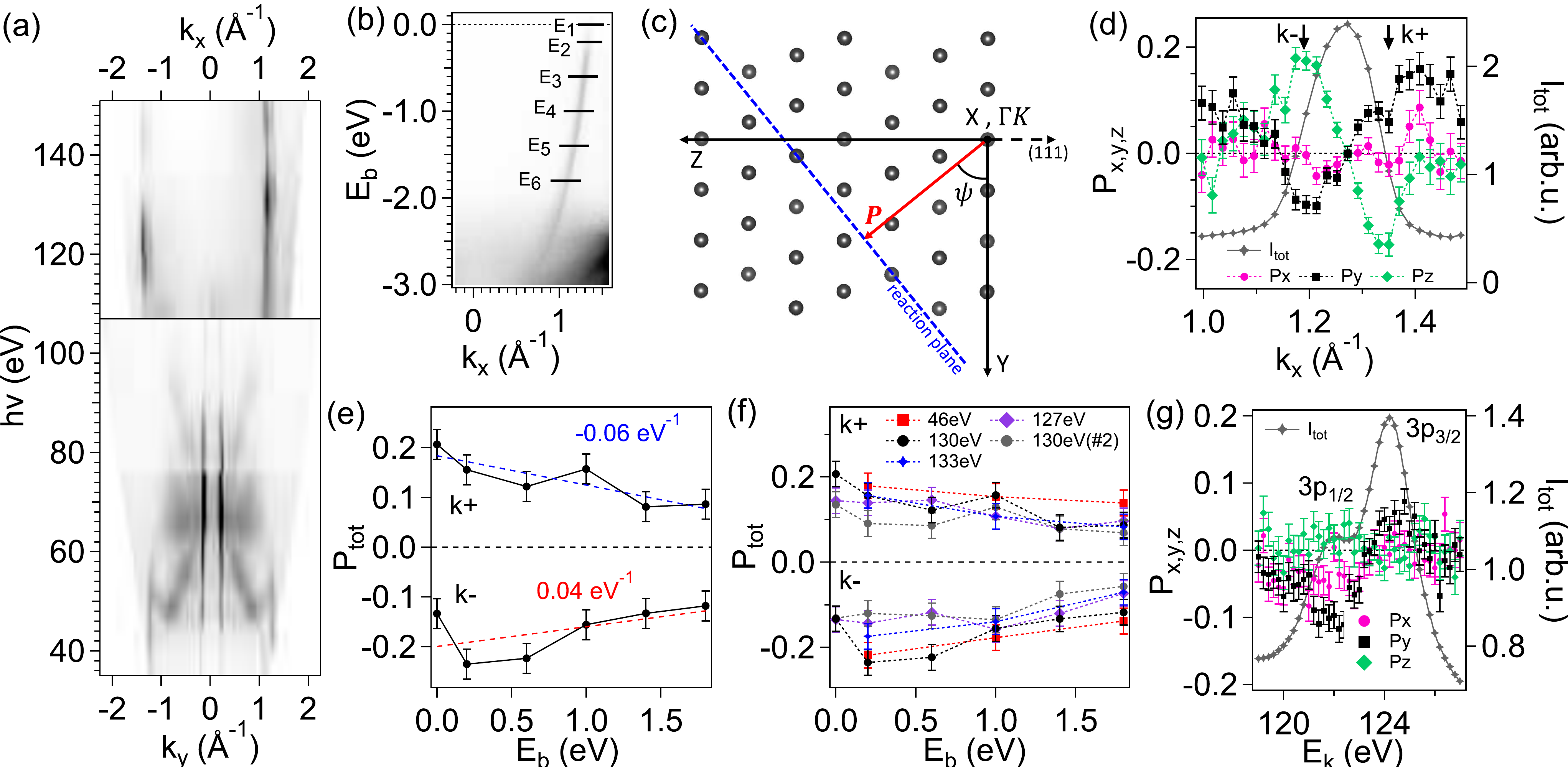}
	\caption{(a) Cu(111) $h\nu$ dispersion for $E_b$ close to the Fermi level; (b) bandmap at $h\nu=130~eV$ with the \textit{sp} bulk CB. Solid lines indicate the $E_b$ where the spin-resolved MDCs have been measured. The actual set was performed in a random sequence in order to prevent any time-related artifact; (c) orientation in space of the measured $\bm{P}$. The reaction plane is tilted by $\psi=51^\circ$ \cite{SOM}; (d) $3$D spin resolved MDC along $x$ measured $0.2~eV$ below the Fermi level with $h\nu=130~eV$ $\pi$ polarized. The total intensity and the three polarization spatial components are shown; (e) plot of $P(E_b)$ for the two spin signals $k^-$ and $k^+$; (f) measurement repeated for different $h\nu$; (g) $3$D spin resolved EDC of the $3p$ core levels with same $E_k$ and $\theta$ as the measurement in (d).}
	\label{fig:data}
\end{figure*}
Our SARPES experiments were performed at the COPHEE endstation \cite{Hoesch:2002, Meier:2009NJP} at the Swiss Light Source. We characterized the spin polarization of the \textit{sp} bulk-derived conduction band (CB) of Cu(111) at room temperature with momentum distribution curves (MDCs) obtained by scanning the angle ${\theta}$ shown in the sketch of the experimental setup in Fig.~\ref{fig:picture}(b). The sample was aligned with the $\Gamma K$ direction along $k_x$ by means of Low Energy Electron Diffraction (LEED) and Fermi surface maps. The sample quality was checked by LEED and by measuring $E_b\approx440~meV$ for the bottom of the surface state \cite{Reinert:2001}, which shows only a Rashba-type spin splitting \cite{SOM} without impurity scattering induced spin interference effects \cite{Dil:2015}.

In order to maximize the counts of the spin-resolved measurements optimal photon energies have been chosen after a $h\nu$ scan, shown in Fig.~\ref{fig:data}(a). Local maxima in intensity were found at $46~eV$ and $130~eV$. A bandmap for $h\nu=130~eV$ is reported in Fig.~\ref{fig:data}(b), which does not show relevant changes in intensity. Solid lines indicate the binding energies where spin-resolved MDCs have been obtained. The CB under consideration displays a nearly-free electron-like dispersion without any hybridization with other bands in a $2~eV$ range from the Fermi level.

In our setup the $E$ field lies in the $xz$ plane with $E_x/E_z\approx0.67$ \cite{SOM} and thus probes both in-plane and out-of-plane orbital components. These in turn are not isotropic as in the simpler case of atomic targets. Because of these symmetries combined with the (111) crystal orientation probed with a low-symmetry non-normal incidence setup the $\bm{\hat{n}}$ direction is not the purely atomic one that would correspond to $y$, but we have to consider the full $3$D vector $\bm{P}$. The measured orientation in space of $\bm{P}$ is shown in Fig.~\ref{fig:data}(c) and can be used to develop a model for the estimate of the term $f(\bm{\Omega})$ \cite{SOM}.

In Fig.~\ref{fig:data}(d) the three spatial components $x$, $y$, $z$ of $\bm{P}$ for the MDC measured $0.2~eV$ below the Fermi level are shown. This clear spin polarization signal has to be generated during the photoemission process since the bulk bands of Cu(111) are spin degenerate in the initial state. We can exclude surface-induced Rashba-like effects \cite{Kimura:2010, Wissing:2013} as the cause for the observed spin polarization because heavier Au(111) shows a polarization with similar magnitude \cite{SOM}. 

The two non-zero components $y,z$ of $\bm{P}$ change sign when crossing the intensity peak maximum, thus resulting in a signal with two peaks (called $k^-$ and $k^+$ in Fig.~\ref{fig:data}). By repeating the measurements at different $E_b$ a plot of $P(E_b)$ was constructed, as shown in Fig.~\ref{fig:data}(e) for both peaks. The slope of their linear fit is the relevant quantity for the determination of $\tau_{\textsc{ews}}$ according to eq.~\eqref{eq:approxtau}, and by applying the model described in ref.~\cite{SOM} we can estimate $\left|\tau_{\textsc{ews}}\right|>26~as$.
The $P(E_b)$ measurement has been repeated for various photon energies as shown in Fig.~\ref{fig:data}(f). To check the robustness the measurement at $h\nu=130~eV$ has been repeated twice. Within the capabilities of the model and of the experimental setup a similar slope is found for $127~eV$ and $133~eV$, but also for the $46~eV$ measurement. This suggests that the time delay considered in our model is not related to the travel time of the electron during the transport to the surface \cite{Neppl:2012, Lemell:2015}.

In order to study the influence of possible additional effects on $P$ we made a survey of spin-resolved energy distribution curves (EDCs) over the $3p$ core levels for angles and kinetic energies corresponding to our CB measurements, as shown in Fig.~\ref{fig:data}(g). Given their localized nature, the core electrons are expected to behave as in atomic photoionization \cite{Roth:1994}. The result of our analysis shows that $\bm{P}$ has only a single peak feature per $3p$ spin-orbit component and does not change for different angles nor kinetic energies, so that diffraction effects through the surface do not play a role \cite{SOM}. 

\begin{figure} [ht]
	\centering
		\includegraphics[width=0.40\textwidth]{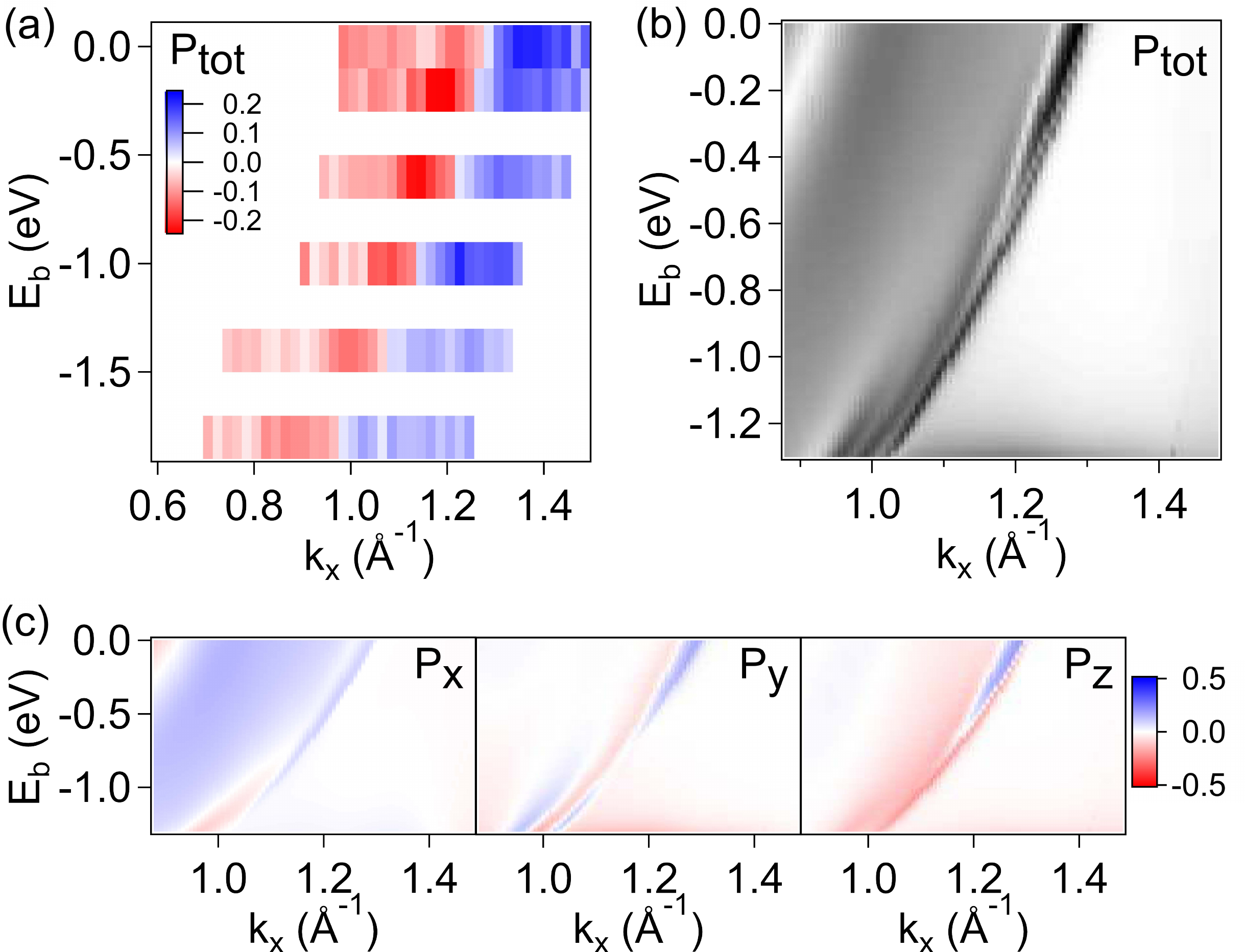}
	\caption{(a) Summary bandmap of spin polarization $P$ from MDCs measured along $k_x$ with $h\nu=130~eV$ $\pi$ polarized; (b) $P$ from KKR calculations performed for similar experimental parameters; (c) the three calculated spatial components of $P$. Their complexity does not permit to unambiguously fix a direction along which (b) could be projected.}
	\label{fig:comparison}
\end{figure}
The two-peaks spin signal of the CB is clearly visible in Fig.~\ref{fig:comparison}(a) where a summary of $P$ from all the MDCs performed with $h\nu=130~eV$ is shown. Crucially, one single band is measured without spin resolution as well established in literature \cite{Courths:1984} and shown in Fig.~\ref{fig:data}(b). In order to understand this critical feature, fully relativistic self consistent multiple-scattering or Korringa-Kohn-Rostoker (KKR) calculations have been performed \cite{Ebert:2011}. In Fig.~\ref{fig:comparison}(b) the evaluated $P$ from ARPES calculations for a semi-infinite system in the framework of the fully relativistic one-step model of photoemission \cite{Braun:1996, Minar:2011b} within its spin-density matrix formulation \cite{Braun:2014} is shown, and its three spatial components are reported in Fig.~\ref{fig:comparison}(c). Also in this case the \textit{sp} band gives rise to two spin signals matching closely to the experimental results, thus excluding any possible artifact in the measurement. This double feature has already been observed, but not discussed, in previous calculations related to self-energy correction studies \cite{Winkelmann:2012}, and will require further investigation.\\

It is important to discuss the nature of the measured time delay since the chronoscopy of photoemission is a fundamental topic in modern physics \cite{Pazourek:2015}. In attosecond-resolved experiments the time delay of a photoelectron beam from a certain state is measured with respect to a different photoelectron beam, which can be from a reference gas system \cite{Locher:2015}, or a different level of the same system \cite{Cavalieri:2007, Neppl:2012}, or the very same state but under different experimental geometry \cite{Lucchini:2015}. Noticeably, the measurement of a finite \textit{relative} time delay $\Delta\tau$ suggests the existence of a finite \textit{absolute} time delay $\tau$ of photoemission for each beam, even though to our knowledge this issue has not been addressed given the impossibility at present to probe absolute times \cite{Neppl:2015}. In this regard, the nature of the time delay indirectly probed by SARPES is not obvious at this stage. It has to be noticed that a time delay of $26~as$ is of the same order of \textit{one cycle} of electromagnetic radiation of $130~eV$, and of the travel time of a $125~eV$ electron along a distance of \textit{one Cu atom radius} in the metallic state.

The spin polarization is produced by the phase shift between the two interfering transitions $T_{1,2}$, which in the case of atomic photoionization correspond to the two final partial waves with orbital quantum number $\ell\pm1$. In solids one should also distinguish between probed in-plane and out-of-plane orbitals \cite{Irmer:1992}, as well as mixed spatial symmetries of the considered state in the double group symmetry representation \cite{Schneider:1989, Yu:1998}, both in the initial and final states. However, the interfering transitions build up the outgoing wave packet so that the net phase shift between them corresponds to the phase of the photoelectron wavefunction. This seems in contrast with the relative time delay probed by time-resolved spectroscopy, and to this extent the two techniques can be considered as complementary. It is also important to underline that the model presented here permits to extract the time information from non-time-resolved calculations, which would be very powerful when performed on systems that are experimentally difficult to probe with time-resolved or spin-resolved ARPES.

In the framework of the one-step model of photoemission it is difficult to tell which process among photon absorption, electron virtual transition and actual photoelectron emission might occur in a finite time. Indeed the influence of the $E$ field on the phase shift is under debate \cite{Nagele:2012, Zhang:2010t, Zhang:2011t} and there might exist a time-threshold for light absorption. A finite decoherence time required by the wavefunction to collapse in the final state might also be considered \cite{Schlosshauer:2005}, and lastly the electron excited above the vacuum level could spend a finite time before reaching the free-particle state. A physical description of the origin of such intrinsic time delay could be a continuous interband coupling mechanism, equivalent for solids of the interchannel coupling in photoionization which leads to finite attosecond time delays \cite{Pabst:2011}. Also the time scale of intrinsic plasmonic satellites might play a role \cite{Lemell:2015}, which could possibly explain the double peak feature of the measured spin polarization in Fig.~\ref{fig:comparison} \cite{Bostwick:2010}. In addition, given the energy-momentum relationship one might be sensitive to spectral variations of the time delay within the band considered \cite{Zhang:2011t}. 

Finally, a note is required about the most common use of SARPES: the study of spin-polarized states. If a spin quantization axis is well defined by the physics of the initial state, interference effects will be concealed, since they contribute only to a small degree of polarization. In fact, whereas a precise quantitative analysis is often impracticable, qualitative results have confirmed many different theoretical predictions. However, it is possible to have a rotation of the spin polarization in half-scattering, and indeed a small rotation of the measured spin polarization compared to theoretical results is quite common in experiments. The development of a more advanced theory of spin-polarized photoemission should take this and other known interference effects \cite{Heinzmann:2012, Meier:2011} into account, together with time delays.

In conclusion, we have derived a semi-quantitative model to access a time delay in photoemission from a dispersive band of a solid by measuring the spin polarization of the photoelectrons. A finite time delay of $|\tau_{EWS}|>26$~as has been found by first experiments on Cu(111) as a model system.\\

We gratefully acknowledge discussions with F.~Da~Pieve. This work was supported by the Swiss National Science Foundation Project No. PP$00$P$2\_144742/1$. This work was supported by the Bundesministerium f\"{u}r Bildung und Forschung (BMBF) under Grant 05K13WMA, the Deutsche Forschunsgsgemeinschaft (DFG) through SPP 1666 and the CENTEM (CZ.1.05/2.1.00/03.0088) and  CENTEM PLUS (LO1402), co-funded by the ERDF as part of the  Ministry of Education of Czech Rep., Youth and Sports OP RDI programme.
\footnotesize
\bibliographystyle{apsrev4-1}
%
\end{document}